\documentclass[12pt]{article}
\usepackage{latexsym,amsfonts,amsmath,amsthm,amssymb,texdraw}

\newcommand{{\bP}}{\bf {P}}

\title{Probability amplitude in quantum like games}

\author{A.A.Grib\thanks{On leave from Alexandre Friedmann Laboratory of Theoretical Physics
St. Petersburg, Russia. e-mail: grib@friedman.usr.lgu.spb.su.} $\; $\thanks{Supported by Profile Mathematical
Modelling in Physics and Cognitive Sciences@Vaxjo University and
EU -network "Quantum Probability and Applications."} , A.Yu.Khrennikov, K.Starkov$^*$.\\
International Centre for Mathematical Modelling\\
in Physics and Cognitive Sciences,\\
University of V\"axj\"o,S-35195 Sweden\\
Email: Andrei.Khrennikov@msi.vxu.se}

\date{}

\begin{document}
\maketitle

\begin{abstract}
    Examples of games between two partners with mixed strategies, calculated
 by the use of the probability  amplitude are given. The first game is
 described by the quantum formalism of spin one half system for which two
 noncommuting observables are measured.

  The second game corresponds to the spin one case.
 Quantum logical orthocomplemented nondistributive lattices for these two
 games are presented. Interference terms for the probability amplitudes are
 analyzed by using so called contextual approach to probability (in the
 von Mises  frequency approach). We underline that our games are not based on 
 using of some microscopic systems. The whole scenario is macroscopic.
\end{abstract}

\section{Introduction}

    It was N.Bohr \cite{Bohr} who said that he believed that the discovery of
  quantum physics really is something more than the discovery of the
  laws of microphysics. He claimed that some aspects (mainly
  complementarity) of quantum mechanics can be manifested in other branches of
  science(biology etc.). However, in spite of the fact that quantum
  formalism proved to be the best description of physical processes
  with molecules, atoms, atomic nuclei and elementary particles, its use
 for other phenomena, to say the least,was not demonstrated explicitly.

    One of the other "fathers" of quantum physics J.von Neumann in his
  paper with G.Birkhoff \cite{Birk} discovered that the specific feature of
  quantum formalism - use of wave functions as probability amplitudes
  is due to the difference of observables of quantum physics from those
  of classical physics. For finite dimensional Hilbert space these
  observables (yes-no questions) form the nondistributive modular
  orthocomplemented lattice. For classical physics this lattice is
  distributive. Later the analysis of foundations of quantum physics and
  its axiomatics in terms of lattices was prolonged by K.Piron,G.Mackey
  and others.

   It became clear that due to nondistributivity of the lattice in case of
   quantum physics it is impossible to define the standard Kolmogorovian
   probability measure and this is the reason for introducing the probability
   amplitude, represented by a vector in Hilbert space.

     A thorough analysis of the difference of probability calculus in
   classical and quantum physics was made by A.Khrennikov in \cite{Kh}, \cite{Kh1} 
   and different aspects of this difference were discussed at numerous conferences on
   foundations of quantum physics in V\"axj\"o\cite{Found}.

    So,following the idea of von Neumann one can say that quantum formalism
   is the description of the new type of random behavior. Chance for this
   behavior is not described by the Kolmogorovian probability.This random
   behavior is repeatable and one can use von Mises analysis of frequencies
   of these or those results when measuring different observables. However,
  due to M.Born rule one must use different probability spaces for
  noncommuting or complementary observables, so that probability measures are
   defined by the wave function differently for these spaces.

    The discovery of Von Neumann shows the way to find other than in microworld
    applications of the quantum formalism. One must look for situations
   when the distributivity law for conjunctions and disjunctions is broken.
   This law however must be broken in such a way that one comes to the
   nondistributive modular orthocomplemented lattices,because only in these
   situations for finite systems there is a possibility to introduce the
 probability amplitudes.In general case one can have nondistributive but not
 modular and orthocomplemented lattices for finite number of observables and
 the description of randomness in such cases is an open question.

  A natural candidate for search of quantum like macroscopic system without
 use of Planck constant is some kind of a game.

   This game must be such organized that disjunction and conjunction defined
  by the rules of the game are not Boolean.

   This means that one can define the rules for conjunction $\bigwedge$ and
disjunction $\bigvee$  experimentally. $C=A\bigvee  B$ if every time A-true follows C-true,
 every time B-true follows C-true.
 $D=A \bigwedge B$ if every time D-true follows A-true,every time D-true follows B-true.
 Then one look for other properties of such C,D in the game considered.
Is C true if and only if A-true or B-true or it can be that C-true also
 in other cases (there is not "only if")?

  Is the rule of distributivity confirmed in our game for all A,B,C?

  To finish the general discussion of the reasons why games are the most
 natural examples of looking for application of the quantum formalism one
 can make a remark that even such a special feature of quantum physics as
 noncommutativity of operators can be found in games.In games one deals with
  "acts" and "acts" very often depend on the order.The typical example is
 that everybody understands the difference arising if one changes the order
 putting on first the shirt and then the suit...

  Examples of games described by vectors in Hilbert space,selfconjugate
 operators as observables realize in some sense N.Bohr's idea of other
 than the microworld situations described by the quantum formalism.
  The first example of such a game, leading to the nondistributive
  orthocomplemented lattice of the spin one half system was given by A.A.Grib and
  G.Parfionov in \cite{GribParf}.

 The payoff matrix for the so called "wise Alice" game was written in terms
 of operators in finite dimensional Hilbert space for spin one half system.
 The average profit of the "wise Alice" was calculated as the expectation
 value of the "payoff operator" in the tensor product of Hilbert spaces for
Alice and her partner Bob.Nash equilibrium points were found for
different
 situations of the "wise Alice" game.

  In this paper we give more explicit than in \cite{GribParf} the rules of the game
"wise Alice". This especially concerns the angles for projections of spin
 operators for Alice and Bob used for different cases of the game.
 Then we shall give some other example of the quantum like game using the
 lattice for spin one system (massive vector meson in quantum physics).
 Differently from our example of "spin one half" wise Alice game here one can
 see the role of nondistributivity more explicitly,because it can be
 expressed in the payoff matrix structure.

  As we said before the peculiar feature of these "quantum-like games" will
be the necessity to use not the Kolmogorovian probability measure but the
probability amplitude.Quantum-like interferences of the alternatives lead to new
rules of calculations of average profits, see [3] for general theory of quantum-like
inteference. That is why when
comparison of the average profit in cases of classical and quantum
like games is made the profit occurs different for these cases.

 So these quantum like games demonstrate the situations where the formalism of
 quantum physics is applied to macroscopic games.Our examples are totally
different from what now is widely discussed in many papers in the
name of "quantum games" \cite{Ek}. All examples with "quantum
coins","quantum gamblers"etc. in this or that way use microobjects
described by quantum physics as some hardware,while in in our
examples everything is totally macroscopic.

 However some results obtained in the cited "quantum game" activity can be
applied to our examples.
 In our examples it is the strategies of Alice and Bob that are described by
 the quantum formalism. Alice, Bob and their acts are totally macroscopic,
 there is no need for the use of Planck constant for them. It is the set of
 frequencies  of their acts which is calculated by use of the quantum
 formalism with wave functions different for Alice and Bob.The optimal
 strategy for both participants is described by the Nash equilibrium and it
is characterized by the special choice  of wave functions for
Alice and Bob  giving the maximal profit for what he/she can get
independently of the acts  of the other partner.

 An interesting feature of our quantum like games is that their description
by the quantum formalism makes necessary use of different probabilistic spaces
 when measuring observables represented by noncommuting operators.Events
(acts)are mathematically described by projectors in Hilbert space. It is
these projectors that form the nondistributive lattice ("quantum logic")
on which the "quantum probability" or the probability amplitude is defined.
 On commuting projectors one defines usual probability measure by the Born
 rule.Typical for the quantum theory interferences of alternatives arise.
  The quantum rule for the average profit takes into account these interferences.
  That is one of the reasons of the difference of the result of
 caculation of this average due to quantum rules and rules of the standard
 probability calculus using Bayesian conditional probabilities.

Finally, we remark that recently there were obtained experimental 
results confirming quantum-like statistical behaviour in psychology, see
\cite{Kh2}; see also \cite{Kh3} on application of the contextual (``quantum-like'')
probability theory in theory of classical spin glasses.

\section{Spin "one half" example of the quantum like game}

  The game "Wise Alice" formulated in the paper\cite{GribParf} is an example of the
 well known game when each of the participants names one of some previously
 considered objects. In the case if the results differ, one of the players
 wins from the other some agreed sum of money.The rules of the game are the
 following.

  1. The participants of our game A and B,call them Alice and Bob,have a
rectangular (in paper \cite{GribParf} it was taken as quadratic!)box in which a ball is
located. Bob puts his ball in one of the corners of the box but does not tell
 his partner which corner. Alice must guess in which corner Bob put his ball.

  2. Alice can ask Bob questions supposing the two-valued answer : "yes" or
  "no". Differently to many usual games the rules of this game are such that:
  
  In the case of a "yes"answer Alice does not receive any money from
  Bob. In the opposite case she asks Bob to pay her some compensation.
  This feature is described by the structure of the payoff matrix.

  3. Differently from other such games \cite{Ow} Bob has the possibility to move the
  ball to any of the adjacent vertices of the rectangle after Alice asks her
  question. This additional condition decisively changes the behavior of Bob
  making him to become active under the influence of questions of Alice. Due
 to the fact that negative answers are not profitable for him he, in all
 possible cases, moves his ball to the convenient adjacent vertex.
  So if Alice asks the question " Are you in the vertex 1? " Bob answers
  "yes" not only in case if he is in 1 but also in cases when he was
  in 2 or 4 due to the possibility to "react" on the question of Alice and moving his
  ball. However, if the ball of Bob was initially in the vertex 3 he cannot
 escape the negative answer notwithstanding to what vertex he moves his ball
 and he fails,letting the ball to be in the same corner.Same rule is valid
 for any vertex. One must pay attention that in this case Alice not only gets
 the profit but also obtains the exact information on the position of the
 ball: honest answer of Bob immediately reveals this position. 
 
 Alice knows about manipulations of Bob.
 
 So negative answers of Bob are valuable for Alice,only getting such answers
 Alice can unambiguously check exclusive positions of Bob. So different
 vertices are incompatible (exclusive) relative to negative answers of Bob.
  This leads to experimentally defined "conjunction"for Alice.

  4. The game is repeated many times,any time Bob putting his ball to some
  corner and Alice guessing this corner.

  5. Due to the rule 3 Alice can draw the graph of the game with four vertices
  and lines connecting them, leading as it was previously done for automata
 in A.A.Grib,R.R.Zapatrin\cite{GribZap} and also the book\cite{GribRod}
 to Hasse diagramm of the spin one half system for which two noncommuting
  spin projections are   measured. The reason for arising of the nondistributive
  lattice represented  by the Hasse diagramm for "events"in "wise Alice" game is
   due to the special
 property of disjunction recognized by Alice experimentally. She sees that
 $1 \bigvee 2 = 2 \bigvee 3 = 3\bigvee 4 = 1\bigvee 3= 2\bigvee 4=I$,
 where $I$ is "always true". Here 1,2, etc. are corresponding questions of Alice.

 It is this unusual
  property of the disjunction that makes Alice consult her friend, who is the
 quantum physicist, for recommendations concerning the frequencies of her
 questions about this or that vertex.

 6. Concerning the frequencies of questions of Alice and Bob's putting the
 ball to this or that vertex one must formulate a special rule interpreting
 not quite clear idea of "preferences" in paper \cite{GribParf}.

  The feature of Hasse diagramm is that it can be represented by projectors
 in Hilbert space nonuniquely by many different ways,making unitary
 transformations of one of the noncommuting operators.

  So let Alice does not know the exact form of the rectangle used by Bob.
 Its diagonals which can be taken of the length 1 in some units can form
 different angles $\theta_B$. Depending on the angle Bob, thinking  that Alice
 will ask him question 1 will put more frequently his ball to that adjacent
  to 1 vertex, the distance of which to 1 is shorter.It can be considered
  as a psychological parameter of the game\cite{GribParf}.

 This angle is fixed for the game. Alice, not knowing the angle $\theta_B$,
 uses the hypothesis that it is some $\theta_A$ and her friend -the physicist
 makes calculations of her average profit using this $\theta_A$ .
 The choice of Alice of $\theta_A$ means that she thinks that the "preference"
of Bob's frequencies will correspond to shortest lines defined by
$\theta_A$.

 So because of fixation of "preferences for Bob's reactions the real free
choice for Bob is the choice between alternative vertices on the diagonals
  of the rectangle 1-3, 2-4. That is why the probabilistic spaces used by
 Alice in her strategy will be different for these two kinds of events.

  Let the payoff matrix of Alice has the structure of a four on four
  matrix $h_{ik}$
  representing payoffs of Alice in each of 16 possible game situations,
 so that one has some positive numbers  a,b,c,d as her payoffs in those
 situations when Bob cannot answer her questions affirmatively. Our game
 is an antagonistic game,so the payoff matrix of Bob is the opposite to
 that of Alice: $(-h_{ik})$.

\begin{center}
 \begin{tabular}{|r|r|r|r|r|}
 \hline
 \multicolumn{1}{|c||}{A/B}
 & \multicolumn{1}{r|}{1}
  & \multicolumn{1}{r|}{2}
  & \multicolumn{1}{r|}{3}
  & \multicolumn{1}{r|}{4}
  \\
  \hline
  1?&0&0&a&0\\
  \hline
  2?&0&0&0&b\\
  \hline
  3?&c&0&0&0\\
  \hline
  4?&0&d&0&0\\
  \hline
\end{tabular}
  \\
~\\
Table 1. The payoff matrix of Alice.
\end{center}

  The main problem of the game theory is to find so called points of
 equilibrium or saddle points - game situations, optimal for all players
 at once. The strategies forming the equilibrium situation are optimal in
 the sense that they provide to each participant the maximum of what he/she
 can get independently of the acts of the other partner. More or less
  rational behavior is possible if there are points of equilibrium defined
 by the structure of the payoff matrix. A simple criterion for the existence
 of equilibrium points is known: the payoff matrix must have the element
 maximal in its column and at the same time minimal in its row. It is easy
 to see that our game does not have such equilibrium point.Nonexistence
 of the saddle point follows from the strict inequality valid for our game
$$
\max_j \min_k h_{jk} < \min_k \max_j h_{jk}
$$

 In spite of the absence of a rational choice at each turn of the game, when
 the game is repeated many times some optimal lines of behavior can be
 found. To find them one must, following von Neumann \cite{vonNeumMorg}, look for the so
 called mixed generalization of the game.In this generalized game the choice
 is made between mixed strategies, i.e. probability distributions of usual
(they are called differently from mixed "pure" strategies) strategies. As
 the criterion for the choice of optimal mixed strategies one takes the
 mathematical expectation value of the payoff which shows how much one can
 win on average by repeating the game many times. In usual classical games
 these expectation values of the payoff are calculated by using the
 Kolmogorovian probability and the optimal strategies for Alice and Bob
 are defined as such probability distributions on the sets of pure strategies
 $x^0=(x^0_1,x^0_2,x^0_3,x^0_4)$ and $y^0=(y^0_1,y^0_2,y^0_3,y^0_4)$
 that for all distributions of x,y the von  Neumann-Nash inequalities are valid:
$$
H_A(x^0, y^0) \geq H_A (x, y^0)
$$
$$
H_B(x^0, y^0) \geq H_B (x^0, y),
$$
where  $H_A, H_B$ -payoff functions of Alice and Bob are the
expectation values of their wins
\begin{equation}
H_A(x, y)=\sum_{j, k=1}^{4} h_{jk} x_j y_k, H_B (x, y)=-\sum_{j, k=1}^4 h_{jk} x_j y_k
\end{equation}

 The combination of strategies, satisfying the von Neumann -Nash inequalities
 is called the situation of equilibrium in Nash's sense. The equilibrium is
 convenient for each player, deviation from it can make the profit smaller.
 In equilibrium situations the strategy of each player is optimal against
the strategy of his(her)partner.
 But as it was shown in \cite{GribParf} the calculation of averages for "wise Alice"
 game, knowing that Bob reacts on her questions must be different from that
 in classical game. Instead of the usual probability measure one must use
 the probability amplitude (the wave function) and calculate probabilities
 for different outcomes by use of the Born's formula.The reason for this
 is the following.
 If one defines the proposition of Alice that Bob's ball is located in
vertex number k (defined by our rule 3) as $\alpha_ k$,then for any
j,k one has
\begin{equation}
\alpha_j \vee \alpha_k=1
\end{equation}
$$
\alpha_j \wedge \alpha_k=0
$$

 Pairs of propositions with the same "parity" $(\alpha_1,\alpha3),(\alpha_2,\alpha4)$ are
 orthocomplemented.
 One has a breaking of the distributivity law. So, for any triple of
 different j,k,l one has the inequality
\begin{equation}
(\alpha_j \vee \alpha_k) \wedge \alpha_e \ne (\alpha_j \wedge \alpha_e) \vee (\alpha_k \wedge \alpha_e)
\end{equation}

 Really, the left side of the inequality is equal to$ \alpha_l$, while the
right side is zero. so the logic of Alice occurs to be a non-distributive
 orthocomplemented lattice described by the Hasse diagramm (Fig.1)

\begin{center}
  \begin{texdraw}
    \drawdim cm
  \move(1 0)\fcir f:0 r:.05
    \move(2 0)\fcir f:0 r:.05
    \move(3 0)\fcir f:0 r:.05
     \move(4 0)\fcir f:0 r:.05
      \move(2.5 1)\fcir f:0 r:.05
   \move(2.5 -1)\fcir f:0 r:.05

                 \move(1 0)\lvec(2.5 1)
                 \move(2 0)\lvec(2.5 1)
                 \move(3 0)\lvec(2.5 1)

                 \move(4 0)\lvec(2.5 1)
               \move(1 0)\lvec(2.5 -1)
                   \move(2 0)\lvec(2.5 -1)
                       \move(3 0)\lvec(2.5 -1)
                           \move(4 0)\lvec(2.5 -1)

    \move(0.8 0)\textref h:C v:B \htext{1}

    \move(1.8 0)\textref h:C v:B \htext{2}

    \move(3.2 0)\textref h:C v:B \htext{3}

    \move(4.2 0)\textref h:C v:B \htext{4}

    \move(2.5 1.3)\textref h:C v:B \htext{I}

    \move(2.5 -1.5)\textref h:C v:B \htext{0}

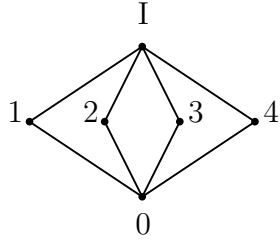
    \move(2.5 -3)\textref h:C v:B \htext{Fig.1 The lattice of Alice's questions and Bob's answers.}
  \end{texdraw}
\end{center}

 On the Hasse diagramm lines going "up" intersect in "disjunction",lines
  going "down"intersect in conjunction.
 The lattice described by our Hasse diagramm is isomorphic to the
 ortholattice of subspaces of the Hilbert space of the quantum system with
 spin one half and the observables $S_x,S_\theta$.For our case it is sufficient
 to take the real (not complex)two dimensional space.So one can draw on the
 plane two pairs of mutually orthogonal direct lines $\{a^1;a^3\},\{a^2;a^4\}$ with the
angle $\theta$  between them coinciding with some angle$\theta_A$ or
$\theta_B$ due to the rule 6 of the game (Fig.2)

\begin{center}
  \begin{texdraw}
    \drawdim cm

                 \move(0 -2)\lvec(0 2)
                 \move(-2 0)\lvec(2 0)
                 \move(-2 1)\lvec(2 -1)

                 \move(-1 -2)\lvec(1 2)

    \move(-0.7 -2)\textref h:C v:B \htext{$a^2$}

    \move(-0.3 0.3)\textref h:C v:B \htext{0}

    \move(-0.3 1.7)\textref h:C v:B \htext{$a^3$}

    \move(2.3 0)\textref h:C v:B \htext{$a^1$}

    \move(2.3 -1)\textref h:C v:B \htext{$a^4$}

    \move(0.7 0.5)\textref h:C v:B \htext{$\theta$}
     \move(3 -2.7)\textref h:C v:B \htext{}

\move (0 0)
\larc r:0.5 sd:0 ed:65

  \end{texdraw}

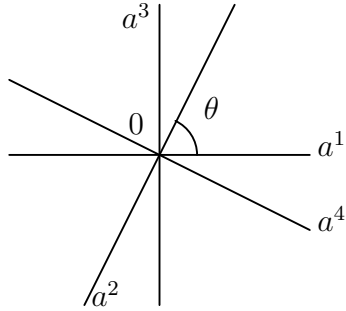
  Fig. 2 Lattice of invariant subspaces of observer $S_x, S_\theta$.
\end{center}

Following the well known constructions of quantum mechanics we
take instead of the sets of pure strategies of Alice and Bob the
pair of two-dimensional Hilbert spaces $H_A,H_B$.Use of Hilbert
space permits us without
 difficulties to realize the nondistributive logic of our players.
 To the predicate $\alpha _k $  put into correspondence the orth projector$\hat{\alpha}_k$. Same
 is done for Bob. Then one writes the self conjugate operator in the space
 $H_A\bigotimes H_B$ which is
the observable of the payoff for Alice

\begin{equation}
\hat H_A=\sum_{j, k=1}^{4} h_{jk} \hat{\alpha}_j \otimes \hat{\beta}_k
\end{equation}

 Let Alice and Bob repeat their game with a ball many times and let us describe
 their behavior by normalized vectors
  $\phi\in H_A$, $\psi\in H_B$,
  so that knowing them one can
 calculate the average according to the standard rules of quantum mechanics as

\begin{equation}
E_{\phi \otimes \psi} \hat H_A=\sum_{j, k=1}^4 h_{jk} < \hat \alpha_j \phi | \phi > <\hat \beta_k \psi, \psi>=
\end{equation}
$$
ap_1 q_3 + c p_3 q_1 + bp_2 q_4 + dp_4 q_2
$$

 Here we take into account our payoff matrix.The operators $\hat{\alpha}_1,\hat{\alpha}_2$ can be written
 as two by  two matrices

\begin{equation}
\hat {\alpha}_1=
\begin{pmatrix}
1&0\\
0&0\\
\end{pmatrix}, \quad
\hat {\alpha}_2 =
\begin{pmatrix}
\cos^2 \theta_A, & \sin \theta_A \cos \theta_A\\
\sin \theta_A \cos \theta_A & \sin^2 \theta_A\\
\end{pmatrix}
\end{equation}

The projectors orthogonal to them are $\hat\alpha_3,\hat\alpha_4$, such that
\begin{equation}
\hat \alpha_1 + \hat \alpha_3 = \hat \alpha_2 + \hat \alpha_4=I
\end{equation}

The wave function $\phi$    can be defined on the plane as some vector with the angle $\alpha$, in general
different from the $\theta_A$.Then the probabilities defined by the Born's rule as
projections on corresponding basic vectors  $a_1,a_2,a_3,a_4$  are
\begin{equation}
p_1=\cos^2\alpha, p_3=\sin^2 \alpha, p_2=\cos^2(\alpha-\theta_A), p_4=\sin^2(\alpha-\theta_A)
\end{equation}
for Alice.
For Bob one has some vector $\psi$   with the angle $\beta$  and $\theta_B$ for $\beta_2$ if $\beta_1$ has the same
form as $\alpha_1$.

\begin{equation}
q_1=\cos^2 \beta, q_3=\sin^2\beta, q_2=\cos^2(\beta-\theta_B), q_4=\sin^2(\beta-\theta_B)
\end{equation}

  The average profit is expressed as some function of $\alpha, \beta$, dependent on
  parameters $\theta_A, \theta_B$:
\begin{equation}
F(\alpha, \beta)=a \cos^2 \alpha \sin^2 \beta + c\sin^2 \alpha \cos^2 \beta +
b\cos^2 (\alpha-\theta_A) \sin^2(\beta-\theta_B) +
\end{equation}
$$
d\sin^2 (\alpha-\theta_A) \cos^2 (\beta-\theta_B)
$$
   This function is defined on the square $[0^0,180^0],[0^0,180^0]$. In paper \cite{GribParf}
   equilibrium points for different values of a,b,c,d and parameters $\theta_A, \theta_B$ were
   found. These points correspond to Nash equilibria.
  Examples with two equilibrium points, one equilibrium point,and no such points
  at all were found.The results strongly depend on the difference of angle parameters
  and values of payoffs in the payoff matrix. Situations with two equilibrium points,
  as well as absence of such a point are obtained for different values of angles
  and not equal a,b,c,d.
   If one takes   $\theta_A=\theta_B=45^0 $     and all a,b,c,d equal to 1 one has the simple solution
    $p_1=1,p_2=0,5,p_3=0,5,p_4=0,5 $ for Alice,
    $q_1=1,q_2=0,5,q_3=0,5,q_4=0,5$ for Bob.
  So the wave functions for Alice and Bob in this case are just eigenfunctions of $\hat\alpha_1$
  The payoff of the "wise Alice" in this case is $E \hat{H}_A=0,5$.
  Due to equivalence of all vertices one can take any other point as "preferable" for
   Alice and Bob,obtaining some other eigenstate of the spin projection operator.The
  "preference" here means that Bob is never putting his ball into this vertex and Alice
   guesses this.The payoff of Alice in all this cases of eigenstates of spin operator
   projections will be the same.
  This can be compared with the   result for the game called in \cite{GribParf} "the foolish Alice"
  who used the standard   probability calculus in the game with the same payoff matrix
  and the same quadrangle   being ignorant about Bob's reactions on her questions.
   Then Nash equilibrium corresponds to equal probability for any vertex and the result
   is  $EH_A=0,25$    which is smaller than    that obtained by her "wise" copy.
    However for the more general case \cite {GribParf}
     $\theta_A=10^0,\theta_B=70^0,a=3,b=3,c=5$
      one has     $\alpha=145,5^0,\beta=149,5^0$
     and  one obtains for the Nash equilibrium in the "wise Alice" game
       $p_1=0,679,p_2=0,509,p_3=0,321,p_4=0,491,
       q_1=0,258,q_2=0;967,q_3=0742,q_4=0,033.$
   Thus, differently from the more or less trivial case,considered before,Bob and Alice
   here don't use the strategy to put the ball in such a manner as neglecting totally
   one of the vertices.Their wave functions now are not just eigenstates of their spin
   operators.

\section{ Spin one quantum like game}

     Now consider the game, described by the graph which coincides with the graph of
   the previous game but with the addition of one isolated point (Fig.3).

\begin{center}
  \begin{texdraw}
    \drawdim cm
                 \move(1 0)\lvec(2 0)
                 \move(2 0)\lvec(2 1)
                 \move(2 1)\lvec(1 1)

                 \move(1  1)\lvec(1 0)
                   \move(1 0)\fcir f:0 r:.05
                     \move(1 1)\fcir f:0 r:.05
                       \move(2 0)\fcir f:0 r:.05
                         \move(2 1)\fcir f:0 r:.05
              \move(3 0.5)\fcir f:0 r:.05

    \move(0.8 -0.2)\textref h:C v:B \htext{\small 4}

    \move(0.8 1)\textref h:C v:B \htext{\small 1}

    \move(2.2 1)\textref h:C v:B \htext{\small 2}

    \move(2.2 -0.2)\textref h:C v:B \htext{\small 3}
    \move(3.2 0.4)\textref h:C v:B \htext{\small 0}

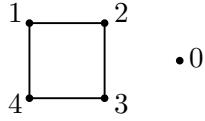
           \move(1 -1)\textref h:C v:B \htext{Fig. 3 The graph of the spin 1 game.}

  \end{texdraw}
\end{center}

   This point is denoted as 0.
   The rules of the game generally are the same as in the previous one.Bob has the same
   possibility to react on Alice's questions by moving his ball to the adjacent vertex and
   only his negative answers are valid for Alice.
    If Bob is in the isolated point 0 he always gives her the "yes" answer which is
    nonprofitable for Alice.
     A new rule will be added by us later making possible for Alice in some cases to ask
     two questions.We shall discuss it when writing the payoff matrix.
    Using the rule of A.A.Grib,R.R.Zapatrin \cite{GribZap} for drawing Hasse diagramms corresponding
   to the graph of automata one obtains for the case of spin one game (Fig.4)

\begin{center}
  \begin{texdraw}
    \drawdim cm
                 \move(1 4)\lvec(3 6)
                 \move(2 4)\lvec(3 6)
                 \move(3 4)\lvec(3 6)
                 \move(4 4)\lvec(3 6)
                 \move(5 4)\lvec(3 6)
                 \move(1 2)\lvec(2 4)
                 \move(2 2)\lvec(3 4)
                 \move(3 2)\lvec(1 4)
                 \move(3 2)\lvec(2 4)
                 \move(3 2)\lvec(4 4)
                 \move(3 2)\lvec(5 4)
                 \move(4 2)\lvec(3 4)
                 \move(4 4)\lvec(5 2)
                 \move(3 4)\lvec(5 2)
                 \move(1 4)\lvec(2 2)
                 \move(1 2)\lvec(3 4)
                 \move(4 2)\lvec(5 4)
                 \move(3 0)\lvec(1 2)
                 \move(3 0)\lvec(2 2)
                 \move(3 0)\lvec(3 2)
                 \move(3 0)\lvec(4 2)
                 \move(3 0)\lvec(5 2)

                   \move(3 6)\fcir f:0 r:.05
                     \move(1 4)\fcir f:0 r:.05
                       \move(2 4)\fcir f:0 r:.05
                         \move(3 4)\fcir f:0 r:.05
                            \move(4 4)\fcir f:0 r:.05
                               \move(5 4)\fcir f:0 r:.05
                                  \move(1 2)\fcir f:0 r:.05
                                     \move(2 2)\fcir f:0 r:.05
                                        \move(3 2)\fcir f:0 r:.05
                                        \move(4 2)\fcir f:0 r:.05
                                        \move(5 2)\fcir f:0 r:.05
                                        \move(3 0)\fcir f:0 r:.05

    \move(0.6 4)\textref h:C v:B \htext{\small $A_8$}

    \move(0.6 1.8)\textref h:C v:B \htext{\small $A_1$}

    \move(1.6 4)\textref h:C v:B \htext{\small $A_6$}

    \move(1.6 1.8)\textref h:C v:B \htext{\small$A_3$}
       \move(2.6 4)\textref h:C v:B \htext{\small$A_5$}
          \move(2.6 1.8)\textref h:C v:B \htext{\small$A_0$}
             \move(3.6 4)\textref h:C v:B \htext{\small$A_9$}
                \move(3.6 1.8)\textref h:C v:B \htext{\small$A_2$}
                   \move(4.5 4)\textref h:C v:B \htext{\small$A_7$}
                      \move(4.5 1.8)\textref h:C v:B \htext{\small$A_4$}
                         \move(3 -0.4)\textref h:C v:B \htext{\small$\emptyset$}
                         \move(3 6.2)\textref h:C v:B \htext{I}

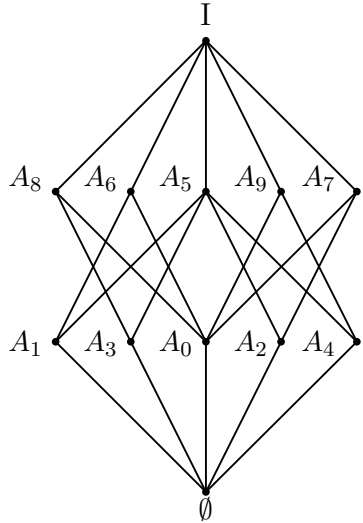
    \move(3 -1)\textref h:C v:B \htext{Fig. 4 Hasse diagramm of the spin 1 game. }

  \end{texdraw}
\end{center}

    It is easy to see that elements $A_1,A_3,A_2,A_4,A_5$ form the same diagramm
   as was considered for the "spin one half" game with the change of  I  on $A_5$.
    But addition of the point 0 leads not only to the appearance of the new "logical
    atom" $A_0$  in our lattice but also to the appearance of the new level composed of
    $A_5,A_6,A_7,A_8,A_9$  which have the meaning of disjunctions
     $A_5=A_1\bigvee A_3\bigvee A_2\bigvee A_4,A_9=A_4\bigvee A_0,A_7=A_2\bigvee A_0$
   Elements of the lattice can be represented by projectors on subspaces of $\mathbb{R}^3$  (Fig.5).

\begin{center}
  \begin{texdraw}
    \drawdim cm
                 \move(0 0)\avec(-1,750 -0.5)
                 \move(0 0)\avec(-1.750 -2)
                 \move(0 0)\avec(0.5 -2)

                 \move(0 0)\avec(2.25 -2)
                  \move(0 0)\avec(0 2)

    \move(-1,750 -0.9)\textref h:C v:B \htext{$A_4$}
    \move(-2 -2)\textref h:C v:B \htext{$A_3$}
    \move(0.8 -2)\textref h:C v:B \htext{$A_2$}
    \move(2.45 -2)\textref h:C v:B \htext{$A_1$}
    \move(0.3 1.8)\textref h:C v:B \htext{$A_0$}

                \move(0.5 -0.8)\textref h:C v:B \htext{$\theta$}

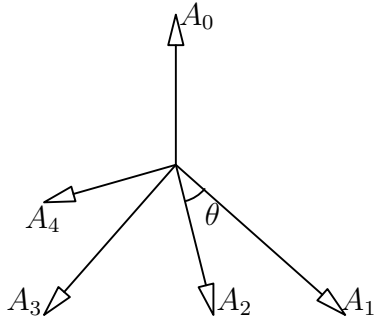
    \move(0 -3)\textref h:C v:B \htext{Fig. 5 Representation of the lattice by supspaces in $R^3$}
   \move (0 0)
   \larc r:0.5 sd:285 ed:320

  \end{texdraw}
\end{center}

    To atoms correspond projectors on lines. $A_1,A_2,A_3,A_4$  are vectors
    in the plane.   One has $A_0\bot A_5,A_2\bot A_4,A_1\bot A_3$.
     The second level is represented by projectors on planes in $\mathbb{R}^3$:
     $A_6$ -- projector on the plane $A_0A_1$,
     $A_7$ -- on $A_0 A_2$,
     $A_8$ -- on $A_0 A_3$,
     $A_9$ -- on $A_0 A_4$.
    One has the orthogonality condition:
     $A_6\bot A_3,A_8\bot A_1,A_7\bot A_4,A_9\bot A_2$.
    Our lattice is the nondistributive modular orthocomplemented lattice.The nondistributivity
   is manifested due to

\begin{equation}
\label{eq11}
A_7 \wedge (A_1 \vee A_4)= A_7 \wedge A_5= A_2 \ne (A_7 \wedge A_1) \vee (A_7 \wedge A_4)=\emptyset
\end{equation}

  This lattice can be represented by selfconjugate operators describing the spin one
system (massive vector meson)for which two noncommuting spin projections $\hat S_z,\hat S_\theta$      are
 measured.One has eigenvalues $S=1,0,-1$. The projector on the zero eigenvalue eigenvector
  is the same in $\hat{S}_z, \hat{S}_\theta$, that is why one has 5 atomic elements in the lattice. All projectors
  can be written as 3 on 3 matrices (see Table 3).

 \begin{table}
  \centering
  \scriptsize
  \begin{gather*}
    A_0 =
    \begin{pmatrix}
      0&0&0\\
      0&0&0\\
      0&0&1
    \end{pmatrix}\qquad
    A_1 =
    \begin{pmatrix}
      1&0&0\\
      0&0&0\\
      0&0&0
    \end{pmatrix}\qquad
    A_2 =
    \begin{pmatrix}
      \cos^2\theta_A & \sin\theta_A\cos \theta_A & 0\\
      \sin\theta_A \cos \theta_A & \sin^2 \theta_A & 0\\
      0&0&0
    \end{pmatrix}\\
    A_3 =
    \begin{pmatrix}
      0&0&0\\
      0&1&0\\
      0&0&0
    \end{pmatrix}\qquad
    A_4 =
    \begin{pmatrix}
      \sin^2\theta_A & -\sin \theta_A \cos \theta_A &0\\
      -\sin\theta_A \cos \theta_A & \cos^2 \theta_A & 0\\
      0&0&0
    \end{pmatrix}\qquad
    A_5 =
    \begin{pmatrix}
      1&0&0\\
      0&1&0\\
      0&0&0
    \end{pmatrix}\\
    A_6 =
    \begin{pmatrix}
      1&0&0\\
      0&0&0\\
      0&0&1
    \end{pmatrix}\qquad
    A_7 =
    \begin{pmatrix}
      \cos^2\theta_A & \sin\theta_A\cos\theta_A &0\\
      \sin\theta_A \cos\theta_A & \sin^2\theta_A & 0\\
      0&0&1
    \end{pmatrix}\qquad
    A_8 =
    \begin{pmatrix}
      0&0&0\\
      0&1&0\\
      0&0&1
    \end{pmatrix}\\
    A_9 =
    \begin{pmatrix}
      \sin^2 \theta_A & -\sin \theta_A \cos \theta_A &0\\
      -\sin \theta_A \cos \theta_A & \cos^2 \theta_A &0\\
      0&0&1
    \end{pmatrix}
  \end{gather*}
\begin{center}
Table 3. Matrices.
 \end{center}

  \end{table}

   These are the well known in quantum mechanics operators of spin for the spin 1 system.
    The operators of observables of Bob are defined analogously, only the angle $\theta_A$ can be
   changed on some $\theta_B$. Now let us discuss the payoff matrix.
    Besides the same rules as were introduced for the "spin one half" quantum like game
   we add a new rule,arising naturally due to the appearance of a new level in comparison
    with the "spin one half " game.
    The rule:Alice can ask not only one question,but two questions in case if the result
    of the first question corresponds to the "disjunction" $\bigvee$. Due to her second question
   she can unambiguously guess where Bob' ball is located.For example Alice asks Bob
    the question:" Are you in 3?"  The answer "no" means he is either in 0 or 1.Then she
   can ask the question:" Are you in 0?" The answer "no" will mean that he is at 1.As we
formulated before only negative answers are valid for Alice.So she
receives from Bob in the considered case of two negative answers
the sum of money $v_1$.   But if she fails she
  receives nothing! If Alice asks the question 0 and the answer is "no",then Bob is in
  $1\bigvee 2\bigvee 3\bigvee 4$ and by guessing correctly this disjunction Alice receives some $u_0$
  for any case. If Alice does not want to risk then receiving the answer "no" on her
question: "Are you in 3?" she doesn't ask the second question and receives some $u_3$.
  Same rule is true for any question.To write the payoff matrix put to the left corner
 the question which in some cases when Alice is going to risk plays the role of the
 second question.
   Putting
\begin{equation}
u_0 < v_1, v_2, v_3, v_4
\end{equation}
\[u_1 < v_0, v_3\]
\[u_2 < v_0, v_4\]
\[u_3 < v_0, v_1\]
\[u_4 < v_0, v_2\]
  where all  $u_i,v_k$ are positive numbers one can define the payoff matrix (see Table 2).

\begin{table}
 \begin{tabular}{|r|r|r|r|r|r|r|}
 \hline
 \multicolumn{1}{|c||}{A/B}
 &\multicolumn{1}{r|}{}
  &\multicolumn{1}{r|}{0}
   &\multicolumn{1}{r|}{1}
    &\multicolumn{1}{r|}{2}
     &\multicolumn{1}{r|}{3}
      &\multicolumn{1}{r|}{4}
      \\
 \hline

  0& &0&$u_0$&$u_0$&$u_0$&$u_0$\\
  \hline
 &1&0&0&0&$v_3$&0\\
\hline
 &2&0&0&0&0&$v_4$\\
 \hline
  &3&0&$v_1$&0&0&0\\
  \hline
   &4&0&0 &$v_2$ &0 &0 \\
   \hline
    1& &$u_1$&0&0&$u_1$&0\\
    \hline
     &0&0&0&0&$v_3$&0\\
     \hline
      &2&$v_0$&0&0&0&0\\
      \hline
       &3&$v_0$&0&0&0&0\\
       \hline
        &4&$v_0$&0&0&0&0\\
\hline
 2&&$u_2$&0&0&0&$u_2$\\
 \hline
  &0&0&0&0&0&$v_4$\\
  \hline
   &1&$v_0$&0&0&0&0\\
   \hline
    &3&$v_0$&0&0&0&0\\
    \hline
     &4&$v_0$&0&0&0&0\\

 \hline
     3 &&$u_3$&$u_3$&0&0&0\\
     \hline
      &0&0&$v_1$&0&0&0\\
      \hline
       &1&$v_0$&0&0&0&0\\
       \hline
        &2&$v_0$&0&0&0&0\\
        \hline
     &4&$v_0$&0&0&0&0\\

\hline
 4&&$u_4$&0&$u_4$&0&0\\
 \hline
  &0&0&0&$v_2$&0&0\\
  \hline
    &1&$v_0$&0&0&0&0\\
    \hline
      &2&$v_0$&0&0&0&0\\
      \hline
        &3&$v_0$&0&0&0&0\\

   \hline
     5&&$v_0$&0&0&0&0\\
     \hline
     6&&0&0&0&$v_3$&0\\
     \hline
     7&&0&0&0&0&$v_4$\\
     \hline
       8&&0&$v_1$&0&0&0\\
  \hline
    9&&0&0&$v_2$&0&0\\
  \hline
\end{tabular}
~\\
~\\
Table 2. The payoff matrix.
\end{table}

  Let us explain again some rules defining the payoff matrix.For example Alice asks the
  question: "Are you in 0?" The answer is "no" ! Then she asks the second question:
  "Are you in 3?" The answer "no " means that Bob is in 1 and he pays $v_1$.    If Alice first
   asks question 1 and gets the "no" answer,then asks 3 and again gets "no" she knows
  that Bob is in 0 and she gets $v_0$.   Same if she is asking 2, 4 and then 3. Questions of
   Alice 5,6,7,8,9 correspond to questions about disjunctions.Negative answers on them make
  possible guessing where Bob is only by this one question.For example "no" answer on 5
  (one must look on the Hasse diagramm Fig.4) means that Bob is in 0 and she gets $v_0$.
  Negative answer on 6 means that he is in 3 and she gets $v_3$   etc.
   The nondistributivity of the lattice is manifested in that,for example
   due to (\ref{eq11})   by asking: "Are you in 0?" and receiving negative answer she concludes that Bob
   is in $A_5=A_1\bigvee A_4$ and then asking:" Are you in 4? and receiving negative answer she concludes that
   he is in $A_2$  and gets the profit $v_2$.   However $A_5$   is also equal to $A_2\bigvee  A_4$    because
 of nonuniqueness of the disjunction, that is why Alice can come to the same conclusion
  without breaking her mind by the nonBoolean nondistributive logic!
   The payoff operator for Alice can be written as
\begin{equation}
\hat H_A=u_0 \hat A_5 \otimes (\hat B_1 + \hat B_2 + \hat B_3 + \hat B_4) +
u_1 \hat A_8 \otimes (\hat B_0 + \hat B_3) +
\end{equation}
\[u_2 \hat A_9 \otimes (\hat B_0 + \hat B_4) + u_3 \hat A_6 \otimes (\hat B_0 + \hat B_1) + u_4 A_7 (\hat B_0 + \hat B_2) + \]
\[v_0 \hat A_0 \otimes \hat B_0 + v_1 \hat A_1 \otimes \hat B_1 + v_2 \hat A_2 \otimes B_2 + v_3 \hat A_3 \otimes \hat B_3 + \]
\[v_4 \hat A_0 \otimes \hat B_4 + v_1 (\hat A_9 \hat A_6 | \otimes \hat B_1 + v_2 (\hat A_9 \hat A_7) \otimes \hat B_2 + v_3 \]
\[(\hat A_5 \hat A_8) \otimes \hat B_3 + v_4 (\hat A_5 \hat A_9) \otimes \hat B_4\]

  The strategies of Alice and Bob are defined as vectors with angles
\begin{equation}
\label{MSI}
\phi=(\cos \alpha_1, \cos \alpha_2, \cos \alpha_3), \psi = (\cos \beta_1, \cos \beta_2, \cos \beta_3)
\end{equation}

   The average profit of Alice is calculated as
\[\alpha_1, \alpha_2, \beta_1, \beta_2 \in [0, \pi], \cos \alpha_3, \cos \beta_3 \geq 0, \]
\[\cos^2 \alpha_1 + \cos^2 \alpha_2 + \cos^2 \alpha_3=1, \cos^2 \beta_1 + \cos^2 \beta_2 + \cos^2 \beta_3=1 \]
\[E\hat H_A=<\phi|\otimes<\psi|\hat H_A|\psi>\otimes|\phi>=\]
\[u_0 p_5 (q_1 + q_2 + q_3 + q_4) + u_1 p_8 (q_0 + q_3) + u_2 p_9 (q_0 + q_4) + \]
\begin{equation}
\label{MSI1}
u_3 p_6 (q_0 + q_1) + u_4 p_7 (q_0 + q_2) + v_0 p_0 q_0 + v_1 p_1 q_1 + 
\end{equation}
\[v_2 p_2 q_2 + v_3 p_3 q_3 + v_4 p_4 q_4\]
    where
\begin{equation}
\label{MSI2}
p_i=<A_i \phi|\phi>, q_i=<B_i \psi|\psi>, 
\end{equation}
\[p_0=\cos^2 \alpha_3, p_1=\cos^2 \alpha_1, p_2=\cos^2\theta_A \cos^2 \alpha_1 + \sin \theta_A \cos\theta_A (\cos^2 \alpha_1 \]
\[+ \cos^2 \alpha_2) + \sin^2 \theta_A \cos^2 \alpha_2, p_3=\cos^2\alpha_2, p_4=1-p_0-p_2,\]
\[p_5=1-p_0, p_6=1-p_3, p_7=1-p_4, p_8=1-p_1, p_9=1-p_2\]
\[q_0=\cos^2 \beta_3, q_1=\cos^2\beta_1, q_2=\cos^2\theta_B\cos^2\beta_1 + \sin\theta_B\cos\theta_B (\cos^2 \beta_1 + \]
\[\cos^2 \beta_2) + \sin\theta_B\cos^2 \beta_2, q_3=\cos^2 \beta_2, q_4=1-q_0-q_2, \]
\[q_5=1-q_0, q_6=1-q_3, q_7=1-q_4, q_8=1-q_1, q_2=1-q_2\]
   For different choices of $\theta_A,\theta_B$ and different $u_i,v_k$ one can obtain different
   Nash equilibria with some fixed values of $\alpha_i,\beta_k$  .

\section{Interference of probability amplitudes in quantum like games}

   Here we shall discuss the following question. Let Alice doesn't know that Bob has
  the facility to move his ball when asked by Alice. Can she get the understanding
  of this facility observing the frequencies of his putting the ball to this or that
  vertex?

   This puts us into analysis of the von Mises frequency probability and the idea of
  the "context" dependence of transformation of probabilities discussed by A.Khrennikov
   in \cite{Kh}.

    In the classical case we have the Bayes' formula
\begin{equation}
\label{LI1}
 p(A=a_i)=p(c=c_1)p(A=a_i/c=c_1) + p(c=c_2)p(A=a_i/c=c_2)
 \end{equation}
    where $A=a_1,a_2$ and  $C=c_1,c_2$ are two dichotomic random variables.In the quantum case
    we have the formula

\begin{equation}
\label{LI2}
p(A=a_i)=p (c=c_1) p(A=a_i/c=c_1) + p(c=c_2)p(A=a_i/c=c_2) 
 \end{equation}
\[\pm 2\sqrt{p (c=c_1) p(A=a_i/c=c_1) p(c=c_2)p(A=a_i/c=c_2)} \cos\theta\]
    where  $\theta$    is some phase.

     For the case of quantum like games (spin one half and spin one cases) this $\theta$  is
     equal to zero and A,C correspond to two spin projections having values +1 or-1, so that
     one has the formula
\begin{equation}
\label{LI3}
p(A=a_i)=p (c=c_1) p(A=a_i/c=c_1) + p(c=c_2) p(A=a_i/c=c_2) 
 \end{equation}
\[\pm 2\sqrt{p (c=c_1) p(A=a_i/c=c_1) p(c=c_2) p(A=a_i/c=c_2)}  \]
      As we discussed before for the general case of Nash equilibrium with some $\theta_A,\theta_B$
      and nonequal values of a,b,c,d for spin one half game the state of Bob is
     some superposition of vectors of the basis for$\hat{S}_{xB}$
\begin{equation}
\label{LI4}
|\psi>=c_1|e_1>+c_3|e_3>
 \end{equation}
         and the probability for definite  $e_1$  is
\begin{equation}
\label{LI5}
|c_1|^2=|<\psi|e_1>|^2
 \end{equation}
        In the basis of $\hat{S}_{\theta B}$  it is
\begin{equation}
\label{LI6}
|\psi>=c_2|e_2>+c_4|e_4>
 \end{equation}
          and
\begin{equation}
\label{LI7}
\vert e_1>=c_\theta|e_2> + \tilde c_\theta|e_4>
 \end{equation}
       so, taking into account  different signs for our coefficients we get
\begin{equation}
\label{LI8}
<\psi|e_1>|^2=|c_2 c_\theta + c_4 \tilde c_\theta|^2=c_2^2 c_\theta^2 + c_4^2 c_\theta^2 
 \end{equation}
\[\pm 2|c_2||c_\theta||c_4|\cdot|\tilde c_\theta|\]
       In our case of real space all the coefficients are real and expressed as
      trigonometric functions of corresponding angles.

     This formula can be understood as the generalization of the Bayes formula and if
     Alice by looking on frequencies can recognize it in Bob's behavior she can understand
      the quantum like type of his game.

       In the game "wise Alice" one has two different complementary contexts - measuring
        $S_x$, meaning Alice asking questions for vertices of the diagonal "1-3" or measuring$S_\theta$,
       Alice asking questions for vertices of the diagonal "2-4".

       But it is important to notice that due to the structure of our game - possibility
       of Bob to change the position of the ball by reacting to the question of Alice-
        we could not "select" for example elements with the property "1-3" without
       disturbing the property "2-4".

        So if one could consider the ensemble of possible game situations for Bob before Alice
      put her questions and call this ensemble $S_0$,then considering questions "1-3"of Alice
      as some filtration leading to ensemble $S_1$,  the ensemble $S_1$   due to Bob's reactions
       will not coincide with $S_0$,i.e.subensemble of nondisturbed Bob's positions.

        Let us analyze from this point of view of "contextually dependent" subensembles
        the simple situation of the "spin one half game" with the quadrangle and equal
       to one all a,b,c,d in the payoff matrix with the Nash equilibrium given by
        $p_1=1,p_3=0,p_2=1/2,p_4=1/2$.

        What does it mean? It means that the " initial" ensemble of Bob was such that he
      never put his ball in 3.As to vertices 2 or 4 he put equally his ball either in 2 or 4.
       Without Alice's questions changing the whole situation one could think of his
       frequencies as if $N_1=2N_2=2N_3=2m$, so that the frequencies for "non perturbed"
       ensembles are
\begin{equation}
\label{LI9}
 \omega_4=\omega_2=\frac{1}{4}, \omega_1=\frac{1}{2}, \omega_3=0
 \end{equation}
        However if Alice makes the "selection" of subensembles asking questions concerning
        ends of diagonals "1-3","2-4" for quadrangle,the whole picture due to Bob's reactions
        will be changed.She will never see the ball in 3 as it was in the initial ensemble,
       but all balls will be moved to 1 when question 1 is aked.This leads to $p_1=1,p_3=0$ situation.

        If questions 2,4 are asked then all the balls will be equally distributed between
       2 and 4! That is the meaning of $p_2=p_4=1/2$.

        So her friend, the quantum physicist,supposing this Nash equilibrium strategy of Bob
        and calculating the average profit for Alice will advice her asking with equal
        frequency questions 2 and 4.Asking question 1 will be unprofitable for Alice because
        Bob will never give her the negative answer. The average profit will be 0,5 as we
        said before. It is easy to see that our rule 6 concerning interpretation of the
        angle for spin projection in terms of "preferences" and corresponding frequencies
         is directly manifested in the situation of eigenstate Nash equilibrium.
         In more general case context dependence will be manifested not so trivially,
         interference terms must be taken into account and recommendations of the
          quantum physicist for Alice will be more sophisticated.

          In conclusion one must make a remark that surely not for all kinds of games
         with "reactions", hidden from one of the partners one necessarily comes to
        the quantum formalism.It is only for special kind of graphs of games (see \cite{GribRod})
        that one comes to Hasse diagramms of the quantum logical nondistributive
        modular orthocomplemented lattice.If the graph is such that the lattice is
        nonmodular or not orthocomplemented one will not have the quantum like structure.
        It is necessary to describe stochasticity in such cases by something different
        either from Kolmogorovian probability or from the quantum probability amplitude.
         Example of such situation is given by the lattice of all topologies for three
         points \cite{GribZ}.

          One of the authors (A. A. G.) is indebted to the International Centre for Mathematical Modelling
in Physics and Cognitive Sciences of Vaxjo University for hospitality.


\begin{thebibliography}{99}

\bibitem{Bohr} N.Bohr. Atomic Physics and Human Knowledge(Science Editions,New York,1961).

\bibitem{Birk} G.Birkhoff and J.von Neumann. The logic of quantum mechanics, Ann.Math. 37,
          823-843 (1936).

\bibitem{Kh} A. Yu. Khrennikov. Linear representations of probabilistic transformations by
          context transitions. J. of Phys. A.34, 9965-9981 (2001).

A. Yu. Khrennikov, Ensemble fluctuations and the origin of quantum probabilistic rule.
J. Math. Phys., {\bf 43}, 789-802 (2002).



\bibitem{Kh1}  A. Yu. Khrennikov, Interpretations of Probability
(VSP Int. Sc. Publishers, Utrecht/Tokyo, 1999).



\bibitem{Found} A. Yu. Khrennikov, Origin of quantum probabilities. Proc. Conf. {\it Foundations of Probability
and Physics,} {\it Quantum Probability and White Noise Analysis}, {\bf 13}, 180-200 (WSP, Singapore, 2001).


A. Yu. Khrennikov, On foundations of quantum theory. Proc. Int. Conf. {\it Quantum Theory: Reconsideration
of Foundations.} Ser. Math. Modelling in Phys., Engin., and Cogn. Sc., vol.2,
163-196 (V\"axj\"o Univ. Press,  2002).



\bibitem{GribParf} A.A.Grib,G.N.Parfionov.Can the game be quantum? Notes of Scient. Sem. Of the
          Petersburg's Branch of the Mathematical Institute of the Russian Academy of
           Sciences, vol.291, p.1-24, St.Petersburg, 2002.

\bibitem{Ek} A.K. Ekert, Phys Rev.Lett.67, p.661,(1999).

\bibitem{Kh2} A. Khrennikov, {\it On cognitive experiments to test quantum-like behaviour 
of mind.} quant-ph/0205092 (2002).


E. Conte, O. Todarello, A. Federici, F. Vitiello, M. Lopane, A. Khrennikov,
A preliminar evidence of quantum-like behaviour in measurements of mental states.
quant-ph/0307201. 


\bibitem{Kh3} A. Yu. Khrennikov, S. V. Kozyrev, Noncommutative probability in classical disordered systems.
{\it Physica A,} {\bf 326}, 456-463 (2003).


\bibitem{Ow} G.Owen. Game theory. W.B.Saunders Company, Philadelphia, London, Toronto.(1968)

\bibitem{GribZap} A.A.Grib, R.R.Zapatrin. Automata, simulating quantum logics. Int. Journ. Theor.
            Phys. 29(2), 113-123 (1990).

\bibitem{GribRod} A.A.Grib, W.A.Rodrigues Jr.Nonlocality in Quantum Physics. Kluwer Academic/
            Plenum Publishers, New York, Boston, Dordrecht, London, Moscow (1999).

\bibitem{vonNeumMorg} J.von Neumann,O.Morgenstern.Theory of Games and Economic Behaviour.
            Princeton.Princeton University Press(1953).

\bibitem{GribZ} A.A.Grib,R.R.Zapatrin. Topologimeter and the problem of physical interpretation
of topology lattice. Int.J.Theor.Phys.,35(3),593-604(1996).

\end{thebibliography}
\end{document}